# Galaxy clusters and the cosmic cycle of baryons across cosmic times

A Whitepaper Submitted to the Decadal Survey Committee

**_Authors_**: R. Giacconi[1], S. Borgani[2], P. Rosati[3], P. Tozzi[4], R. Gilli[5], S. Murray[6], M. Paolillo[7], G. Pareschi[8], G. Tagliaferri[8], A. Ptak[1], A. Vikhlinin[6], K. Flanagan[13], M. Weisskopf[9], A. Bignamini[4], M. Donahue[10], A. Evrard[11], W. Forman[6], C. Jones[6], S. Molendi[12], J. Santos[4] & G. Voit[10]

1. Dept. of Physics and Astronomy, The Johns Hopkins University, Baltimore MD
2. Dept. of Astronomy, University of Trieste, Trieste, Italy
3. European Southern Observatory (ESO), Garching bei Muenchen, Germany
4. INAF-Astronomical Observatory of Trieste, Italy
5. INAF-Astronomical Observatory of Bologna, Italy
6. Harvard-Smithsonian Center for Astrophysics, Cambridge MA
7. Dept. of Physical Sciences, Universita' Federico II, Napoli, Italy
8. INAF-Astronomical Observatory of Brera, Milano, Italy
9. NASA Marshall Space Flight Center, Huntsville AL
10. Physics and Astronomy Dept., Michigan State University, East Lansing MI
11. Dept. of Physics and Astronomy, University of Michigan, Ann Arbor MI
12. INAF-Istituto di Astrofisica Spaziale, Milano, Italy
13. Space Telescope Science Institute, Baltimore MD

## _Science Frontier Panels_
**Primary Panel:** Galaxies across Cosmic Time (GCT)
**Secondary panel:** Cosmology and Fundamental Physics (GFP)

**_Project emphasized_: The Wide-Field X-Ray Telescope (WFXT)**
http://wfxt.pha.jhu.edu

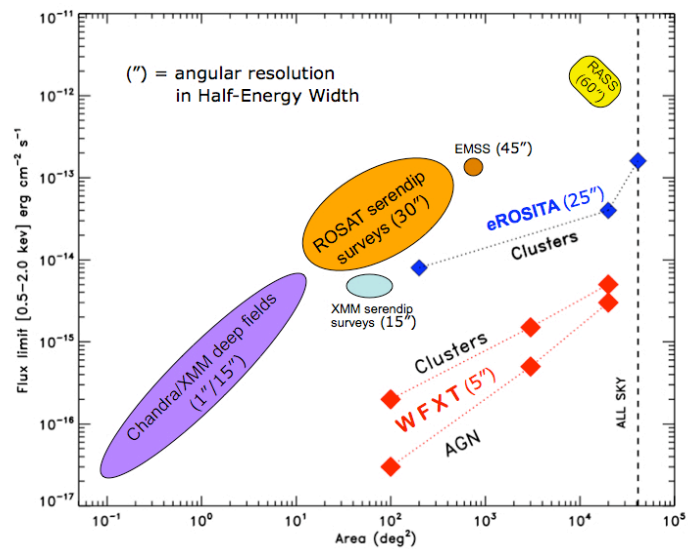

**Executive summary.** We discuss the central role played by the X-ray study of hot baryons within galaxy clusters to reconstruct the assembly of cosmic structures and to trace the past history of star formation and accretion onto supermassive Black Holes (BHs). We shortly review the progress in this field contributed by the current generation of X-ray telescopes. Then, we focus on the outstanding scientific questions that have been opened by observations carried out in the last years and that represent the legacy of Chandra and XMM: (a) When and how is entropy injected into the inter-galactic medium (IGM)? (b) What is the history of metal enrichment of the IGM? (c) What physical mechanisms determine the presence of cool cores in galaxy clusters? (d) How is the appearance of proto-clusters at $z\sim2$ related to the peak of star formation activity and BH accretion? We show that the most efficient observational strategy to address these questions is to carry out a large-area X-ray survey, reaching a sensitivity comparable to that of deep Chandra and XMM pointings, but extending over several thousands of square degrees. A similar survey can only be carried out with a Wide-Field X-ray Telescope (WFXT), which combines a high survey speed with a sharp PSF across the entire FoV. We emphasize the important synergies that WFXT will have with a number of future ground-based and space telescopes, covering from the radio to the X-ray bands. Finally, we discuss the immense legacy value that such a mission will have for extragalactic astronomy at large.

## 1. Galaxy clusters: the legacy of Chandra and XMM

Galaxy clusters represent the place where astrophysics and cosmology meet: while their overall internal dynamics is dominated by gravity, the astrophysical processes taking place on galactic scale leave observable imprints on the diffuse hot gas trapped within their potential wells[1]. Understanding in detail the relative role played by different astrophysical phenomena in determining this cosmic cycle of baryons, and its relationship with the process of galaxy formation, is one of the most important challenges of modern cosmology.

Clusters of galaxies represent the end result of the collapse of density fluctuations over comoving scales of about 10 Mpc. For this reason, they mark the transition between two distinct dynamical regimes. On scales roughly above 10 Mpc, the evolution of the structure of the universe is mainly driven by gravity and the evolution still feels the imprint of the cosmological initial conditions. At scales below 1 Mpc the physics of baryons plays an important role in addition to gravity, thus making physical modeling far more complex. In the current paradigm of structure formation, clusters form via a hierarchical sequence of gravitational mergers and accretion of smaller systems. Within these small halos gas efficiently cools, forms stars and accretes onto supermassive black holes (SMBHs), living in massive galaxies, already at high redshift. While the star formation peaks at $z\sim2$-3, the intergalactic gas is heated to high, X-ray emitting temperatures by adiabatic compression and shocks, and settles in hydrostatic equilibrium within the cluster potential well, only at relatively low redshift, $z<2$. The process of cooling and formation of stars and SMBHs can then result in energetic feedback due to supernovae or AGN, which inject substantial amounts of heat into the intergalactic medium (IGM) and spread heavy elements throughout the forming clusters.

Thanks to the high density and temperature reached by the gas within them, galaxy clusters mark the only regions where thermo- and chemo-dynamical properties of the IGM can be studied in detail at $z<1$ from X-ray emission, and directly connected to the optical/near-IR properties of the galaxy population. A remarkable leap forward in the quality of X-ray observations of clusters took place with the advent of the Chandra and XMM-Newton satellites. Thanks to their unprecedented sensitivity (and angular resolution in case of Chandra), they led to a number of fundamental discoveries concerning nearby, $z<0.3$, clusters. For instance:

**(a)** The lack of strong emission lines at soft X-ray energies in the core regions placed strong limits on the amount of gas that can cool to low temperatures[2], thus challenging the classical "cooling flow" model[3];

**(b)** Temperature profiles have been unambiguously observed to decline outside the core regions and out to the largest radii sampled so far (~$R_{500}$[1]), while they gently decline toward the cluster center in relaxed systems[4];

**(c)** The level of gas entropy at $R_{500}$ is in excess of what explainable by the action of supersonic accretion shocks[5], while it is unexpectedly low in the innermost regions of relaxed clusters[6];

**(d)** The intra-cluster medium (ICM) is not uniformly enriched in metals, instead metallicity profiles are observed to have a spike in the central regions, associated to the presence of the brightest cluster galaxy (BCG), while declining at least out to $0.3R_{500}$ [7].

While these observations shed new light on our understanding of the physical properties of the low-redshift intergalactic medium, (IGM), they opened at the same time at least as many questions.

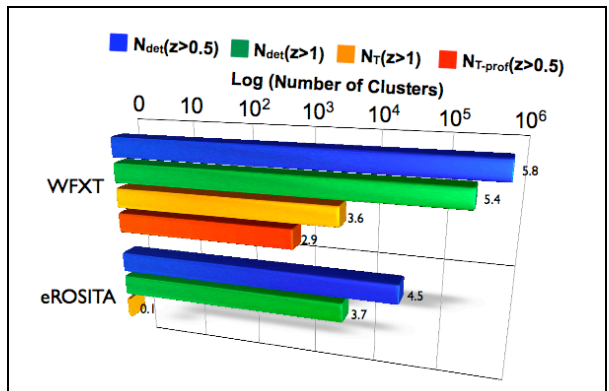

**Figure 1.** The cluster discovery space for a high-sensitivity, 5-year long Wide Field X-ray Telescope survey, compared with the planned eROSITA mission (2013). Bars indicate the estimated number of clusters detected at z>0.5 and z>1 (blue and green respectively, assuming that at least 50 source counts are needed), number of clusters at z>1 for which an accurate measurement of the temperature can be obtained (>1500 counts required; orange) and those at z>0.5 for which temperature profiles and metallicity can be recovered (>15,000 counts required; red). Note that a logarithmic scale is used.

## 2. Key open questions

**Q1. When and how is entropy injected into the IGM?** The standard explanation for the excess of entropy observed out to $R_{500}$ is that some energetic phenomena should have heated the ICM over the cluster life-time[1]. Models based on the so-called pre-heating (i.e. diffuse entropy injection before the bulk of the mass is accreted into the cluster halos) have been proposed as an explanation[8]. However, these models predict quite large isentropic cores, which are not observed. Furthermore, observations of the inter-galactic medium (IGM), from observations of z~2 absorption systems in high-resolution optical spectra of distant QSOs, demonstrate that any pre-heating should take place only in high-density regions[9]. An alternative scenario is that ICM heating takes place at relatively low redshift, within an already assembled deep potential well. In this case, the natural expectation is that the same heating agent, presumably the central AGN, should be responsible for both establishing the cool core and increasing the entropy out to ~ 1 Mpc scale, although it is not clear how AGN feedback can be distributed within such a large portion of the cluster volume.

Reconstructing the timing and pattern of entropy injection in the ICM has far reaching implications in tracing the past history of star formation and black hole (BH) accretion. While we expect that the two above scenarios leave distinct signatures on the evolution of the ICM entropy structure, available data from XMM and Chandra are too sparse to adequately understand this evolution.

---

1 We indicate with $R_{\Delta c}$ the cluster-centric radius encompassing an average overdensity $\Delta_c$ times the critical cosmic density. For reference, $\Delta_c$ =200 is close to the virial overdensity while $\Delta_c$ =500 corresponds to ~half the virial radius for a concordance $\Lambda$CDM model.

**Q2. What is the history of metal enrichment of the IGM?** This question is inextricably linked to the previous one on the history of IGM heating. Measurements of the metal content of the ICM provide direct information on the past history of star formation and processes (e.g., galactic ejecta powered by SN and AGN, ram-pressure stripping of merging galaxies, stochastic gas motions, etc.) that are expected to displace metal-enriched gas from star forming regions[10]. So far, X-ray observations have provided valuable information on the pattern of enrichment only at low-redshift (z<0.3)[11]. Profiles of the Fe abundance have been measured for nearby systems[12]. However, these results are limited to rather small radii, <0.3$R_{500}$, while the level of enrichment at larger radii should be quite sensitive to the timing of metal production and to the mechanism of metal transport. Furthermore, profiles of chemical abundances for other elements, such as O, Si, and Mg, are much more uncertain[12]. Tracing the relative abundances of different chemical species, which are synthetized in different proportions by different stellar populations, is crucial to infer the relative role played by different SN types and to establish the time-scale over which the ICM enrichment took place. The situation is even more uncertain at z>0.3. Although analyses based (mainly) on the Chandra archive show indications for an increase of the ICM enrichment since z~1[13], basically no information is available on the metallicity profiles and on abundance of elements other than Fe. To improve upon this situation, one needs *(a)* to push to larger radii the study of the distribution in the ICM of different chemical species in nearby clusters and *(b)* to measure profiles of the Fe abundance for hundreds of clusters at z>0.5.

**Q3. What physical mechanisms determine the presence of cool cores in galaxy clusters?** XMM and Chandra unambiguously demonstrated that the rate of gas cooling in cluster cores is unexpectedly low. This result removed the previously claimed inconsistency between high mass-deposition rate inferred from X-ray imaging of cluster cores and the low level of star formation observed in central cluster galaxies. However, such a low cooling rate requires that some sort of energy feedback must heat the ICM so as to exactly balance radiative losses. AGN are generally considered as the natural solution to this problem[14]. However, no consensus has been reached so far on the relative role played by AGN and by mergers in determining the occurrence of cool cores in galaxy clusters. Since merging activity and galactic nuclear activity are both expected to evolve with redshift, measurements of the occurrence of cool cores in distant clusters are necessary to address this issue. Although attempts have been pursued to characterize the evolution of the fraction of cool cores using Chandra data[15], no definite conclusion has been reached on the evolution of the fraction of cool core clusters.

**Q4. How is the appearance of proto-clusters related to the peak of star formation activity and BH accretion?** Massive galaxies in today's clusters show only very modest ongoing star formation: they harbor a super-massive black hole usually living in a quiescent accretion mode and experience only "dry" mergers with much smaller galaxies. The situation should be radically different at z~2. This is the epoch when proto-BCGs are expected to be assembling through violent mergers between actively star-bursting galaxies, moving within a rapidly evolving potential well. These proto-cluster regions accrete a large amount of gas that is suddenly heated to high temperature by mechanical shocks and, for the first time, starts radiating in X-rays. At the same time, BHs hosted within merging galaxies are expected to coalesce and their accretion disks to be destabilized by the intense dynamical activity, thereby triggering a powerful release of feedback energy. Evidence for such forming proto-clusters has been obtained by optical observations of a strong galaxy overdensity region, the so-called Spiderweb complex, surrounding a previously identified powerful radio galaxy at z~2[16]. Cosmological simulations lend support to the expectation that similar structures trace the progenitors of massive cluster seen locally, and predict that this structure should already contain dense IGM, emitting in X-rays with $L_X$~$10^{44}$ cgs in the [0.5-2] keV band, with a temperature of several keV and enriched in metal at a level comparable to nearby clusters[17]. As of today, no unambiguous detection of X-ray emitting gas permeating this region has been obtained[18]. While the detection of such a hot diffuse gas may be just at the limit of the capability of current X-ray telescopes, characterizing its physical properties (temperature and metallicity) is far beyond the reach of Chandra and XMM. Expectations based on the standard LCDM model with WMAP-5y cosmological parameters suggest that several hundreds of these systems should be present at z>2 over the whole sky.

## 3. The need for a sensitive X-ray survey of galaxy clusters

The best observational strategy to address the above questions is to carry out high-sensitivity surveys in the soft X-ray band (i.e. [0.5-4] keV) over a large area of the sky. The three most important characteristics that an X-ray survey telescope must have for this purpose are the following:

**i)** A large field-of-view (FOV) combined with a large collecting area (A), i.e. with unprecedented grasp (FOV x A), which measures the survey speed.

**ii)** High angular resolution *across the entire FoV*, to distinguish low surface brightness extended sources, to miminize point-source contamination and to carry out high-quality imaging and spectroscopy in the central regions of distant (out to z~1) clusters.

**iii)** Suitable choice of the orbit of the satellite (e.g. low equatorial) to minimize the particle background, so as to take full advantage of the instrument sensitivity and of the high-quality Point Spread Function (PSF).

The design of an X-ray telescope with such characteristics is beyond the scope of this White Paper. However, for our purpose, here it is enough to state that the requirements of FOV=1 sq.deg., A=0.5-1 $m^2$ at 1-2 keV and a PSF of about 5" (half power diameter), approximately constant over the entire FOV, can be met by the polynomial profiles for X-ray mirrors[19] with a realistic technological development over a time-scale of about 5 years. Such a mission, Wide Field X-ray Telescope (WFXT), has been proposed and described by Murray et al.[22] Carrying out large area surveys with a Wide Field X-ray Telescope (WFXT), having such a large grasp and high resolution, represents the most effective observational strategy to address the aforementioned questions.

In Figure 1, we show a comparison for the yields of clusters expected from five years of operation of WFXT, compared with the expectations for the planned German-led mission eROSITA[^2]. Besides the huge number of clusters that the WFXT will detect at large redshift, this demonstrates that measurements of the physical properties of the ICM will be available for a large number of them.

In Figure 2 we show a simulation of a 1 sq.deg. of X-ray sky as observed by a WFXT in only 13 ksec. This image elucidates the power of combining large collecting area and sharp PSF for the characterization of a large number of galaxy clusters. As an example, in the top left cut-out we show how a moderately luminous cluster at z=1.6 is detected as an extended source with ~300 counts. Such distant clusters will be routinely detected in each WFXT field-of-view.

Also shown in the central left panel is the relation between true redshift and redshift recovered from the detection of the Fe line in the X-ray spectra of the brightest clusters in this field. This will open the way to an unprecedented, entirely X-ray selected, cluster redshift survey, which will include tens of thousands systems at 0.5<z<~1.5, thereby avoiding extremely time consuming optical spectroscopy.

**Answer to Q1.** A large number of clusters with ~$10^4$ counts will increase by orders of magnitude the statistics of a handful of clusters at z>0.5 for which detailed entropy profiles have been measured. For these distant clusters the ICM properties will be measured with a quality

[^2]: Based on the eROSITA Mission Definition Document available at http://www.mpe.mpg.de/erosita/MDD-6.pdf

comparable to what Chandra and XMM provided so far at z<0.2 for a smaller number of objects. The measurement of ICM profiles out to z~1 and beyond will allow us to trace the interplay between IGM and galaxy population along 2/3 of the cosmological past light-cone. Furthermore, the low background and the possibility of resolving out the contribution of point sources will also allow us to measure such profiles out to $R_{200}$ and beyond for bright galaxy clusters at z<0.2.

**Answer to Q2.** Iron metallicity profiles will be measured for virtually all the clusters for which a temperature profile is obtained, although with ~2 times larger statistical errors. A very accurate measurement of the global Fe metallicity will be obtained for tens of thousands of clusters out to z~1.5. For all the clusters of this sample, thermo-dynamical and chemical properties of the ICM will be characterized with unprecedented precision.

**Answer to Q3.** The high-quality PSF will allow one to resolve the core region of distant clusters (a cool-core of 50 kpc radius will subtend an angle of ~6" at z=1). The yield of hundreds of clusters at z>0.5 for which more than $10^4$ counts will be available, will allow us to accurately measure the evolution of the occurrence of cool cores and how this is related to the cluster dynamical state.

**Answer to Q4.** The study of proto-clusters at z~2 is still unexplored territory. For this reason, it is difficult to make predictions on how many of these structures will be observed. By extrapolating our present knowledge of the relation between mass and X-ray luminosity, we expect to detect several hundreds of such objects over the whole sky. For the brightest of these clusters, it will even be possible to measure their redshift through X-ray spectroscopy with deeper follow-up exposures. At z~2 the inverse Compton scattering of relativistic electrons, injected by AGN in core regions, off the CMB photons is much more effective than at low-z in producing a hard X-ray excess, thanks to the higher CMB temperature. At high redshift the K-correction also increases the inverse Compton emission relative to thermal bremsstrahlung at softer observed photon energies. The bottom left panel of Fig. 2 shows a deep 400 ksec WFXT pointing on a proto-cluster, formed in a cosmological hydrodynamic simulation[17]. The large number (~15,000) of photons detected for this object will allow one: (a) to catch "in fieri" the pristine ICM enrichment; (b) to see in action the combined effect of strong mergers and intense nuclear activity within a forming cluster; (c) to discern the thermal and non-thermal emission from X-ray spectroscopy and infer the early contribution of cosmic rays in pressurizing the ICM. The comparison with expectations for eROSITA (Fig. 1) demonstrates that the goal of measuring physical properties of the ICM out to z~1 and beyond can only be accomplished by a survey with the area and sensitivity achievable with WFXT. In fact, WFXT constitutes a two orders of magnitude improvement with respect to eROSITA (similar to the area-sensitivity enhancement that eROSITA will give with respect to the ROSAT All-Sky Survey), with in addition a 6 times better angular resolution. In summary, WFXT will not be just a highly efficient cluster-counting machine, aimed at detecting ~$10^6$ objects out to their formation redshift. Its unique added value is that it will characterize the physical properties for a good fraction of these clusters and, therefore, calibrate them as robust tools for cosmological applications. Besides the astrophysical study of the cosmic cycle of baryons, on which this White Paper is focused, an obvious application will be the reconstruction of the growth rate of cosmic structures and, from this, the determination of the underlying cosmological model, without the need of carrying out any follow-up observations with other large-area X-ray telescopes (aspect discussed in a White Paper by A. Vikhlinin et al.).

## 4. Synergies with other non X-ray surveys & legacy value

Addressing the outstanding questions outlined above will greatly benefit from a coordinated multi-wavelength activity between WFXT, future space missions and ground-based facilities.
 **Synergies with ground-based optical surveys.** The identification and characterization of the galaxy populations hosted by the half million X-ray luminous clusters at z>0.5, unveiled by WFXT, will be an essential process to obtain a comprehensive and self-consistent picture of the cosmic cycle of baryons in their hot and cold phase, by tracing the evolution of their underlying stellar populations and star formation histories. Fortunately, deep optical coverage of the large survey areas will be provided by the next generation of wide-field ground-based facilities, currently under development and scheduled for routine operations within the next few years. Specifically LSST and Pan-Starrs will have the necessary depth and multi-band (0.4-1μ) imaging to enable immediate identification of a large number of clusters, also providing accurate photometric redshifts.

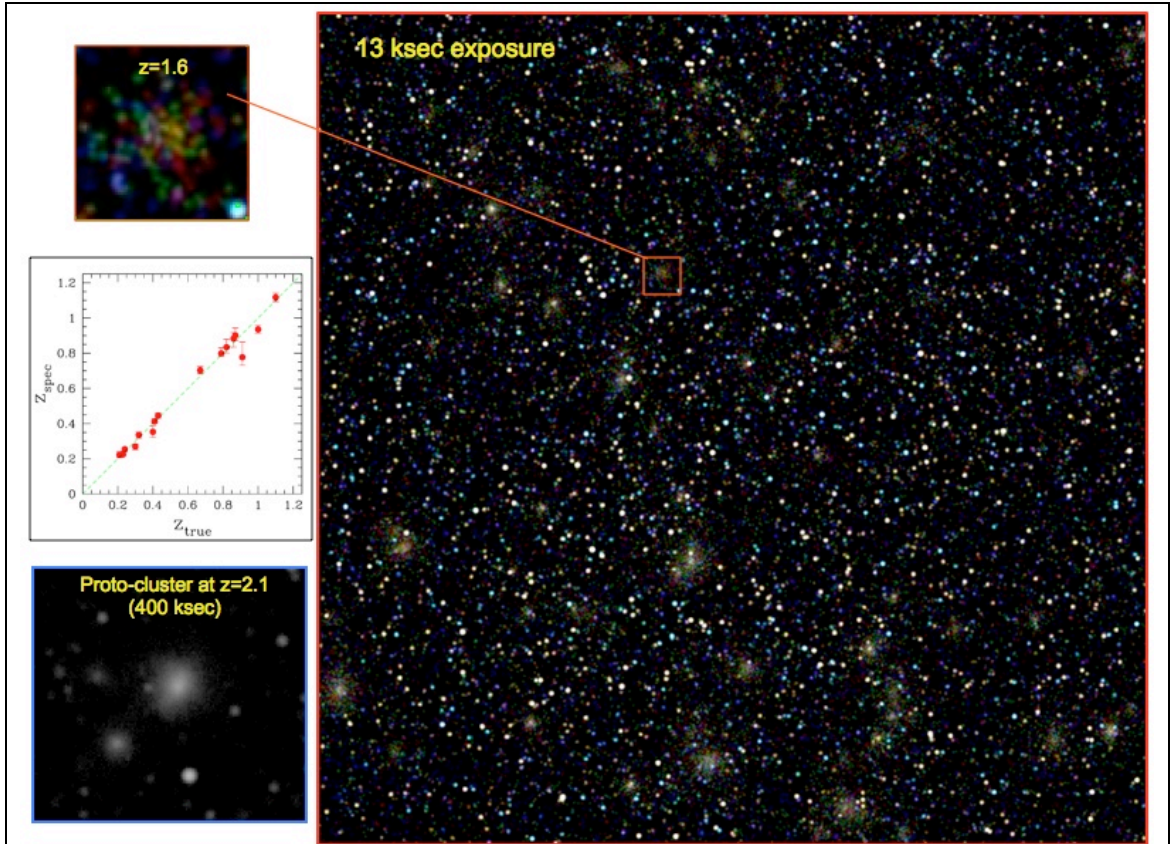

**Figure 2.** Simulation of a WFXT image (FoV=1 deg$^2$) with 13 ksec exposure . Red, green and blue colors denote X-rays of increasing energy, from 0.5 to 7 keV. The field has been populated with clusters obtained by "cloning" at high z a set of clusters observed with Chandra at low z, so as to reproduce the evolution of the X-ray luminosity function predicted in a WMAP-5y cosmology. The AGN population has been modelled according to the predictions of a population synthesis model for the X-ray background. The flux limit for cluster detection is ~10$^{-15}$ cgs in the [0.5-2] keV band. The field includes 69 clusters above this flux limit, with 19 lying at z>1. *Top left*: a 3 arcmin cut-out of a cluster at z=1.6, with $L_X$=8×10$^{43}$ cgs in the [0.5-2] keV band. *Central left*: the relationship between true redshift and redshift recovered from the X-ray spectrum (using the Iron K$\alpha$ line) for the 17 brightest clusters within this field (with at least 500 counts). *Bottom left*: 10 arcmin cut-out of a proto-cluster at z=2.1, extracted from a cosmological hydrodynamic simulation, as observed with an exposure of 400 ksec. This is the progenitor of today's massive cluster, predicted to have $L_X$=4×10$^{44}$ cgs and detected with ~15,000 counts.

**Synergies with JDEM.** The combination of WFXT with NASA and ESA dark energy (DE) missions currently under development (JDEM and Euclid) will allow unprecedented studies of clusters at z>1, as well as proto-clusters at z~2, by providing their spectroscopic confirmation

(very challenging and time consuming from the ground) and a full characterization of member galaxies with high resolution rest frame optical imaging (0.5-1.7μ). DE missions are also designed to reconstruct the DM mass distribution via weak lensing tomographic techniques. This will allow direct lensing mass determination of thousands of massive clusters out to z≈1. Their comparison with X-ray derived masses will yield the much heralded cluster mass calibration and control of systematics for high-precision cosmological applications.

**Synergies with Sunyaev-Zeldovich surveys.** The Atacama Cosmology Telescope (ACT) and the South Pole Telescope (SPT) have recently opened a new era of Sunyaev-Zeldovich (SZ) cluster search[20]. Next generation large single-dish mm telescopes, such as the Caltech-Cornell Atacama Telescope (CCAT) will have enough sensitivity and angular resolution to carry out large-area SZ surveys, providing at the same time spatially resolved SZ imaging for moderately distant massive clusters. Taking advantage of the different dependence of the SZ and X-ray signals on gas density and temperature, their combination will provide a reconstruction of temperature and mass profiles, independent of X-ray spectroscopy[21]. This will offer further independent means of calibrating mass measurements of clusters.

**Legacy value.** With its *unprecedented grasp and angular resolution*, WFXT will be an outstanding source of interesting targets for follow-up studies with facilities such as JWST, ALMA, the next generation of giant (30-40m) ground-based telescopes, and X-ray observatories (i.e., IXO and Gen-X). For example, a combined study of X-ray luminous proto-cluster regions with ALMA, will test whether a phase of vigorous star formation (sub-mm bright galaxies) coexist with a BH accretion phase. WFXT will also provide targets for future X-ray missions with large collecting area. Indeed, follow-up pointed observations with IXO will allow the study of metallicity and entropy structure of the ICM in the most distant systems at z~2.

In general, the synergy with next generation multi-wavelength deep wide-area surveys and with high sensitivity instruments for pointed observations will unleash the full potential of WFXT in addressing a number of outstanding scientific questions for the next decade and will consolidate its immense legacy value.